# Review of Hybrid Load Balancing Algorithms in Cloud Computing Environment

[1],*Chukwuneke, Chiamaka Ijeoma; [2]Prof. Inyiama, Hyacinth C.; [3]Amaefule, Samuel;
[4]Onyesolu, Moses Okechukwu; [5]Asogwa, Doris Chinedu,
[1,3,4,5]Department of Computer Science, Nnamdi Azikiwe University, Awka, Nigeria.
[2]Department of Electronics and Computer Engineering, Nnamdi Azikiwe Unversity, Awka, Nigeria.

*Abstract:* In cloud computing environment, load balancing is a key issue which is required to distribute the dynamic workload over multiple machines to make certain that no single machine is overloaded. In recent research, many organizations lose significant part of their revenues in handling the requests given by the clients over the web servers i.e. unable to balance the load for web servers which results in loss of data, delay in time and increased costs. Various static and dynamic algorithms have been proposed and implemented in the past but this have not been fully efficient for load balancing. This gave room to hybrid algorithms. Hybrid methods inherit the properties from both static and dynamic load balancing techniques and attemptsat overcoming the limitation of both algorithms. This paper is a study of various hybrid load balancing algorithms in cloud computing environment.

## I. INTRODUCTION

Cloud computing is an emerging technology and has attracted a lot of attention in both commercial and academic spheres. Cloud computing has moved computing and data away from desktop and portable PCs into large data centers (Sharma & Banga, 2013). In cloud computing, users do not know where the infrastructures are located. The users only use the services through the cloud infrastructure paradigm and pay for the requested services (Armbrust et al., 2010). Virtualization is a key enabling technology for cloud computing environments, which makes it possible to run multiple operating systems and multiple applications on the same hardware at the same time, so as to provide services by a virtual unit (Armbrust et al., 2010). The National Institute of Standards and Technology's (NIST) define cloud computing as " a model for enabling ubiquitous, convenient, on-demand network access to a shared pool of configurable computing resources (e.g., networks, servers, storage, applications and services) that can be rapidly provisioned and released with minimal management effort or service provider interaction (Mell & Grance, 2011).

In cloud computing, the biggest challenge is how to handle and service the millions of requests that are arriving very frequently from end users efficiently and correctly. Thereby, the need for load balancing. Load balancing is one of the most important issues in cloud computing to improve the performance. Load balancing is a process of distributing load. The load is distributed on individual nodes to maximize throughput, and to minimize the response time. It also removes a condition in which some of the nodes are heavily loaded while some others are light (Ramana, Subramanyam &AnandaRao, 2011)

In general load balancing algorithms are classified as static or dynamic and centralized or distributed (Abubakar et al.,2004). A comprehensive study was carried out, where the comparisons are conducted among static, dynamic, and hybrid algorithms. The results show that the hybrid algorithms give efficient results since they inherit the benefits of other algorithms, and avoid drawbacks (Siham, 2017).The objective of this paper review is to unveil the existing hybrid load balancing techniques in cloud computing which are more efficient and will take in to account the overall network loads, energy efficiency with better quality of service satisfaction.

## II. LOAD BALANCING

Load balancing is a computer networking method to distribute workload across multiple computers or a computer cluster, network links, central processing units, disk drives, or other resources, to achieve optimal resource utilization, maximize throughput, minimize response time, and avoid overload (Kaur &Bansal,2013) Load balancing appears to be the major challenge in cloud computing, due to heterogeneous nature of cloud environment where resource pool is on increase (Gabi, Ismail & Zainal, 2015). The load balancer accepts multiple requests from the client and distributing each of them across multiple computers or network devices based on how busy the computer or network device is. Load balancing helps to prevent a server or network device from getting overwhelmed with requests and helps to distribute the work (Gabi, Ismail & Zainal (2015). Providing an efficient load balancing in cloud computing enables efficient resource utilization, achieved higher user satisfaction and also prioritizes users by applying appropriate scheduling criteria (Katyal & Mishra,2012).

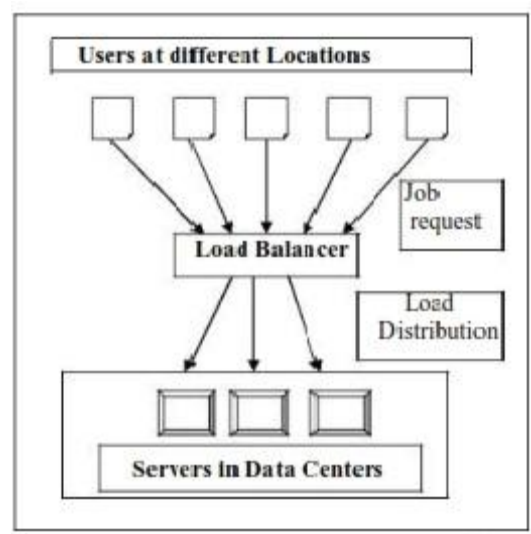

Figure 1:Load Balancing Technique (Karthika, Kanakambal & Balasubramaniam, 2015)

*A. Goals of Load Balancing*

The goals of load balancing are:

  a. To improve the performance substantially

b. To have a backup plan in case the system fails even partially
c. To maintain the system stability
d. To accommodate future modification in the system

*B. Load Balancing Algorithm*

The purpose of load balancing algorithm is to improve the performance by redistributing the workload among available server nodes (Rajguru & Apte,2012). Load balancing techniques are normally used for balancing the workload of distributed systems. Load balancing algorithms are broadly classified into two types (Zahra & Khalid, 2016) namely:

i.) Static load balancing
ii.) Dynamic load balancing.

**i. Static Load Balancing Algorithms**

In this approach (Hamadah, 2017), the load balancing is performed by previous information with regard to the system. Next, depending on the performance workload is distributed withoutconsidering the current state of the node. Once the load isallocated to the node, it cannot be transferred to another one.The nodes perform their work and send back the results totheir clients. Example: Round Robin,Min-Min,Max-Min, etc.

**ii. Dynamic Load Balancing Algorithms**

These algorithms monitor changes on the system workload and redistribute the entire works. This algorithm is usually composed of three strategies, which comprise: the transfer strategy, the location strategy, and the information strategy.The transfer strategy decide which tasks are eligible to be transferred to other nodes for processing. The location strategy nominates a remote node in order to execute a transferred task.The information strategy represents the information center forthe Load Balancing algorithm (Beniwal & Garg, 2014). In fact, it is responsible for providing the location and the transferred strategies to each node. The Dynamic Load Balancing algorithm can be achieved based on three ways: non-distributed, distributed, or semi distributed methods (Hamadah, 2017). In the non-distributed method, there is one node (centralized) that receives all requests and distributes them to the servers. In the distributed method, all nodes are shared with the distribution of the requests. As for the semidistributed method, the nodes are divided up into a group of clusters, where each cluster works as a central node in order to distribute the requests, and all clusters are responsible for the load balancing distribution. Example Least Load,Throttled load Balancing Algorithm, Ant Colony Optimization, Join Idle-Queue etc.

*C. Hybrid load balancing algorithms*

These algorithms are achieved and are proposed to eliminate the drawbacks of dynamic and static load balancing methods, and they are being used to aggregate the benefits and merits of static and dynamic algorithms in order to design a new one (Milani & Navimipour,2016). In fact, this implies that the combinations of the benefits of two or more existing algorithms either dynamic or static algorithms are able to present a new one.

### III. RELATED WORK

Javanmardi et al. (2014) proposed a hybrid job scheduling algorithm using genetic algorithm and fuzzy theory. The proposed algorithm assigns jobs to resources without considering the job length and resources capacities. Genetic algorithm was used as the basis of the approach. This was modified with the aid of fuzzy theory to reduce the iteration of producing the population. Two types of chromosomes with different QOS parameters were defined; then the fitness value of all the chromosomes for the mentioned two types were obtained. The new approach with the use of fuzzy theory modified the standard genetic algorithm and improved system performance in terms of execution cost to 45% and total execution time to 50%.This algorithm was implemented on CloudSim and the results showed that this hybrid approach outperformed other algorithms.

Bhowmik et al. (2016) proposed an efficient load balancing approach in a cloud computing platform.This work proposes a priority based virtual machine (VM) arrangement and load balancing using round robin scheduling. The VM is arranged by comparing the vital parameters which involve bandwidth, RAM, and MIPS (million instructions per second). The new algorithm is based on round robin for load balancing and ensures that jobs are not scheduled according to the VM priority and hence no VM is inundate or remains inactive for a longer period of time. This new method eliminates the process of excess burden on any individual VM and enhances the resource utilization. Cloudsim was used as a simulation tool to simulate the algorithms. The experiment results clearly indicated that the proposed method task scheduling consumes lesser time than conventional Round Robin Algorithm.

Wang et al., (2011) proposed a two-phase scheduling load balancing algorithm (OLB+LBMM). This algorithm combines Opportunistic Load Balancing (OLB) and Load Balance Min-Min (LBMM) scheduling algorithms. OLB scheduling algorithm keeps every node in working state to achieve the goal of load balance and LBMM scheduling algorithm is utilized to minimize the execution of time of each task on the node thereby minimizing the overall completion time. This algorithm works to enhance the utilization of resources and enhances the work efficiency.

Tasquia et al.(2012) proposed a modified task scheduling algorithm based on the concept of beelife algorithm (BLA) and greedy algorithm to get an optimal service in hybrid cloud. Bee life algorithm are residential for task scheduling while the greedy method will select randomly forone data center. In modified job scheduling the jobs are processed in the queue using non-preemptive priority queue. First the tasks enter to the BLA algorithm and to greedy method to get optimal solution. The make span can be reduced by this hybrid algorithm.

Liu, Luo, Zhang, Zhang and Li (2013) proposed a multi-objective genetic algorithm (MOGA)-based scheduling algorithm. This algorithm combines both random and greedy initialization methods. In genetic algorithm the fitness is calculated by energy consumption and profits of the service providers. The best fitness is selected and stored in pareto. Then selection operation can be done by two strategies: elitism and crowding. After the selection process crossover operation can be done by two individuals. Mutation swaps their position to generate new individuals This MO-GA is used to minimize the energy consumption and to maximize the profit of service providers.

Parsa and Entezari-Maleki (2009) proposed a new algorithm RASA (Resource Aware Scheduling Algorithm): A new task scheduling algorithm in grid environment. The algorithm was built through a comprehensive study and analysis of two well-known task scheduling algorithms, Min-min and Max-min. RASA uses the advantages of both algorithms and covers their disadvantages. According to this algorithm at first the scheduler allocates the resource to tasks according to number of available resources. The available resource is added then it chooses Min-Min algorithm procedure first. It allocates the resource to tasks in round one-time smaller tasks in next round to large tasks, vice-versa. The simulation result showed that this algorithm outperformed the two conventional algorithms compared with it.

Rajput, (2016) proposed a genetic based improved load balanced min-min task scheduling algorithm for load balancing in cloud computing. This workproposed an improved load balanced min-min (ILBMM) algorithm using genetic algorithm (GA) in order to minimize the makespan and increase the utilization of resource. Firstly,the execution time of task on the virtual machine is calculated. With that, the minimum or maximum time of task on virtual machine (VM) is known.Genetic based approach was then applied on million instructions (MI) of task and million instructions per second (MIPS) of VM for better execution. Here we used a crossover, mutation and fitness function of GA.The implementation of proposed algorithm was completed using CloudSim simulator and the simulation outcome demonstrated that the proposed algorithm outperformed the current algorithm on same objectives.

Lin, Liu and Wu (2011) proposed an energy- efficient virtual machine provision algorithm for cloud system. These authors proposed the combination of dynamic Round-Robin and First-Fit to form a hybrid algorithm. The probability distribution (e.g., a normal distribution) is followed and the number of incoming virtual machines is assumed as a function for time. Hybrid algorithm used the virtual machine's incoming rate for scheduling of virtual machines. The First- Fit was used by the Hybrid method during rush hours to completely utilize the computing power of physical machines, and then used the dynamic Round-Robin for the consolidation of the virtual machines and thus reduced the consumption of the energy in non-rush hours.

## IV. EXISTING HYBRID LOAD BALANCING ALGORITHM

a. Younis, Halees and Radi, (2015) proposed a hybrid load balancing algorithm in heterogeneous cloud environment. In this research, a hybrid algorithm that takes advantages of both random and greedy algorithms was proposed. The algorithm adopted the characteristics of randomization and greedy to make an efficient load balancing algorithm that covered their disadvantages. The algorithm considered the current resource information and the CPU capacity factor to achieve the objectives. The hybrid algorithm consists of two main steps:

1. In the first step Virtual Machines (VMs) is distribute over hosts accordingto the host qualifications. The largest number of VMs islocated at the most qualified host depending on theHosts' CPU capacity.

2. In the second step the algorithm used a new index tableto record the current loads for each VMwhich isused to check the current VM loads at each iteration. The algorithm reads the value of VM load from the indextable; when the data center receives a request from theusers, it sends the request to the hybrid load balancer.The hybrid algorithm will select k nodes (VM) randomly, and then it will choose the current load for each selected VM. Then it will choose a VM that have least VM current loads and return the VM id to datacenter. The data center will assign the load to the selected VM and update the value of selected VM in the index table of current loads. Finally, when the VM finishes processing the request, it will inform the datacenter to update its current load value. The hybrid algorithm has been evaluated and compared with other algorithms using cloud Analyst simulator. The experiment results showed that the algorithm developed by Younis, Halees and Radi, (2015) improved the average response time and average processing time compared with other algorithms.

### HLBA Algorithm

Input: new request

Output: The VM id that selected to assign the load.

0. Initialize, Cl_Table(0..n-1) ← 0 At start all VM's have zero allocation., K← m, VM_id ←-1 , VMids()=-1,i← 0, currCount ← 0, minCount ← Max_Value, TempVMid ← 1;
1. Parses VM_List() to LoadBalancer:
2. For i← 0 to k  //Select VM randomly
3.    TempVMid ← random(VM_List()).
4. VM_id ← TempVMid
5. If vm_id Exist in Cl_Table(VM_id) then
6. currCount ← Cl_Table(VM_id)
7. Else
8. currCount ← 0
9.    VMids() ← (VM_id, currCount).
10. End for
11. TempVMid ← -1
12. currCount ← 0
13. For i ← 0 to k
14. TempVMid ← i
15. currCount ← VMids(TempVMid)
16. If currCount<minCount then
17. minCount= currCount
18.    VM_id ← TempVMid
19.    End if
20. End for
21. Cl_Table(VM_id) ← Cl_Table(VM_id)+ 1

b. Subalakshmi & Malarvizhi, (2017) proposed an enhanced hybrid approach for load balancing algorithms in cloud computing. Enhanced hybrid approach is the advancement of hybrid algorithm which contains both Throttled and Equally Spread Current Execution algorithm. Enhanced hybrid algorithm maintains an index list of VM allocation status as well as list to count the allocated request. The allocated request list is compared with theVMs index list. If VMs index list is greater than allocated request list it means that VMs are available to take request else request has been queued until VM isbeen available. If the VM has been queued, it has towait in the queue itself. So new host has been created using host create function. In case of availability of VM, the jobs are allocated to that particular VM. And both the index list and hash list are updated. The job in queue need not wait for long time for the virtual machine to become available. Enhanced hybrid approach load balancing algorithm is proposed and implemented in cloud computing environment using CloudSim toolkit, in java language. By analyzing the parameters in graphs and tables we came to know that the overall response time, data processing time is relatively minimized as well as data transfer cost is reduced.

### TESCE Algorithm

Step 1: Initialize the allocation status of all VMs as AVAILABLE in the state list of VM.

Step 2: Initialize Hash map without any entries.

Step 3: A new request has been sent to the DataCenterController (DCC).

Step 4: For next allocation new load balancer has been queried by the DataCenterController.

- Step 5: If Hash map list size less than VM state list size then allocate VM.
- Step 6: Else call host create function using which new physical machine configuration has.
- Step 7: After finishing the processing of request by VM Data Center Controller receives cloudlet response and VM de-allocation is noticed.
- Step 8: The status of VM in VMs state list and Hash map list is updated by the load balancer.

Singhal, Shah and kalantri, (2011) proposed a load balancing algorithm over a distributed cloud network.

This study proposed the use of a hybrid task scheduling algorithm which combines two commonly used scheduling methods, the MM (Min-Min) and OLB (Opportunistic Load Balancing) to create our hybrid Balanced Load Min-Min (BLMM) algorithm. In this study, a three-level hierarchical framework is used to reach load balance and decrease execution time for each node. The third level is the service node used to execute subtask. The second level is the service manager used to divide the task into some logical independent subtasks. The first level is the request manager that assigned the task to a suitable service manager node. BLMM considers the execution time of each subtask on each service node. Each subtask will figure out the execution time on different service nodes (N11, N12…) via agent. According to the information gathering by agent, each service manager chooses the service node of shortest execution time to execute different subtasks and records it into the Min-time array. Finally, the Min-time array of each subtask is recorded, that is a set of minimal execution time on certain service nodes. Meanwhile, service manager chooses the service node from Min-time array. This means the $a^{th}$ subtask on service node 'g' is performed first.Therefore, the subtask 'a' will be distributed to service node 'g'. Since the subtask 'a' has been distributed to service node 'g' to be performed, thesubtask 'a' will be deleted from subtask queue. The Min-time array will be rearranged, and the service node 'g' is put on the last one of Min-timearray. In this study, an integrated scheduling algorithm is provided that combines OLB and Min-Min. According to the properties of the proposed integrated scheduling algorithm, the load balance and the execution time of each node is considered.

### BLMM Algorithm

- Step 1: According to the requirement of subtask 'Ti' to choose the minimal execution time service node from N service nodes, and to form a Min-timeservice node set, where 'Ti' is the total number of subtasks, N is the total number of nodes inside the executable subtask service nodes set.
- Step 2: Choose the service node 'g' from Min-timeset in which has the shortest execution time,where 'g' is the identifier number of node.
- Step 3: Assign subtask 'a' to service node 'g'.
- Step 4: Remove the complete executed subtask 'a' from the needed to be executed task set, where 'a'is represented the identifier of subtask.
- Step 5: Rearrange the Min-time array, and put the service node 'g' at the last.
- Step 6: Repeat Step 1 to Step 5, until all subtasks have been executed.

d. Domanal and Reddy, (2015) proposed a hybrid approach using divide-and conquer and throttled algorithm for load balancing. This algorithm combines the methodology of Divide and-conquer and Throttled Algorithm (DCBT) which schedules the incoming requests to available VMs efficiently and ensures that there is no starvation of the requests. This hybrid approach consists of two algorithms: Pass I and Pass II. According to this algorithm requests from different clients are provided to the available Request Handlers (RH) and Virtual machines (VM). In the initial step, this algorithm checks for the availability of RH's and VM's and divides the requests accordingly using divide and conquer approach. In the next step, the incoming requests are assigned to the different RH's and VM's. Load Balancer keeps track of the current status of each RH or VM and verifies that the current request should only be assigned to the RH and VM which has not been used recently. This algorithm ensures that the load is distributed in an optimized way and no resource is idle thus leading to maximum resource utilization and minimum execution time thus leading to high performance. The algorithms for Pass I and Pass II are as follows:

### DCBT Algorithm: Pass I

- Step 1: Find the number of tasks in queue for a period 't'
- Step 2: Find the number of available RH in a real time distributed system.
- Step 3: Tasks are divided continuously by the number of available RH in a real time distributed system.
- Step 4: After the division, check remainder == number of RH then, assign tasks to the RH for execution
- Step 5: Step 1 to 4 iterates continuously in coordination with pass II until it finishes the execution.

### DCBT Algorithm: Pass II

- Step 1: Initially Load Balancer assigns the tasks to the available Request Handlers or Virtual Machines.
- Step 2: For the next assignment, Load Balancer checks for the availability of RH.
- Step 3: Next task is allocated to available RH or VM, iff
  1. RH or VM should be free.
  2. Assigned RH or VM should be used for the previous assignment.
- Step 4: Step 1 to 3 are repeated in coordination with Pass I until all the tasks are executed completely.

e. Hung, Wang andHu, (2012) proposed an efficient load balancing algorithm for cloud computing network

In this work, we propose an efficient load balance algorithm, named Load Balance Max-Min and Max algorithm (LB3M). LB3M, which combines minimum completion time and load balancing strategies. LB3M assign tasks to computing nodes according to their resource capability. From the case study, LB3M achieves better load balancing and minimum completion time for completing all tasks than other algorithms such as MM and LBMM

### LB3M Algorithm

- Step 1: It is to calculate the average completion time of each task for all nodes, respectively.
- Step 2: It is to find the task that has the maximum average completion time.
- Step 3: It is to find the unassigned node that has the minimum completion time less than the maximum average completion time for the task selected in Step 2. Then, this task is dispatched to the selected node for computation.
- Step 4: If there is no unassigned node can be selected in Step 2, all nodes including unassigned and assigned nodes should be reevaluated. The minimum completion time of an assigned node is the sum of minimum completion time of assigned task on this node and the minimum completion time of the current task. The minimum completion time of an unassigned node is the current minimum completion time for the task. It is to find the unassigned node or assigned node that has the minimum completion time less than the maximum average completion time for the task selected in Step 2. Then, this task is dispatched to the selected node for computation.

Step 5: Repeat Step 2 to Step 4, until all tasks have been completed totally.

f. Gao and Wu, (2015) proposed a dynamic load balancing strategy for cloud computing with ant colony optimization. This work presented a novel approach on load balancing via ant colony optimization (ACO), for balancing the workload in a cloud computing platform dynamically. Two strategies, forward-backward ant mechanism and max-min rules, are introduced to quickly find out the candidate nodes for load balancing. We formulate pheromone initialization and pheromone update according to physical resources under the cloud computing environment, including pheromone evaporation, incentive, and punishment rules, *etc*. Combined with task execution prediction, we definethe moving probability of ants in two ways, that is, whether the forward ant meets the backward ant, or not, in the neighbor node, with the aim of accelerating searching processes. Simulations illustrate that the proposed strategy can not only provide dynamic load balancing for cloud computing with less searching time but can also get high network performance under medium and heavily loaded contexts.

### ACO Algorithm

1. Beginning of proposed algorithm
2. Initialize pheromone for slave nodes;
3. Get-job-from-user ($job_n$) for master node;
4. Job-divides-into-tasks ($job_n$) by master node;
5. for (i = 0 to $n_t$) {//$n_t$is the distribution number of the tasks
6. Distribute-tasks-to-slaves ($task_i$);
7. If-there-are-overload/underload-nodes () {
8. Generate-forward-ant ();
9. Compute-moving-probability ();
10. Move to next node;
11. if(node-is-candidate)
12. Generate-backward-ant ();
13. Start-timer-for-backward-ant ($timer_{na}$);
14. Update-pheromone-by-forward-ant ();
15. if ($timer_{na}$> 0)
16. Update-pheromone-by-backward-ant ();
17. if(task-in-slave-successful)
18. Increase- pheromone;
19. if(task-in-slave-failed)
20. Decrease- pheromone;
21. }
22. if(satisfy-load-balancing) {
23. Do-load-balancing ();
24. continue;
25. }
26. else if(need-new-tasks)
27. Go to 3;
28. }
29. End of algorithm

g. Kumar and Verma, (2012) proposed a hybrid scheduling algorithm which is an improved version of genetic algorithm applied to scheduling algorithm**.** The work used genetic scheduling algorithm to reach at optimal conditions used for scheduling. The authors used Min-Min, Max-Min and completion time parameters. The Min-Min and Max-Min scheduling methods are merged in standard Genetic Algorithm. This algorithm schedule multiple jobs on multiple machines in an efficient manner such that the jobs take the minimum time for completion, however, they can further be improved by adding hardware constrains. It also does not consider other parameters like as storage capacity, early execution time, workload type, job size. The idea for generating the initial population randomly in genetic algorithm is replaced by Min-Min and Max-Min which increases the chances of produces better child. The experiment results show improved genetic algorithm is much better compare to the standard genetic algorithm.

### MM Algorithm

1. Begin
2. Find out the solution by Min-Min and Max-Min
3. Initialize population by the result of Step 2
4. Evaluate each candidate
5. Repeat Until (termination condition occurs)
    a. Select parents
    b. Recombine pairs of parents
    c. Mutate the resulting offsprings
    d. Evaluate new candidate
    e. Select individuals for next generation
6. End

h. Saini and Bisht, (2015) proposed a hybrid algorithm for load balancing. This work focuses on two commonly used load balancing techniques: Round-Robin algorithm (Static Load balancing) and Least Load algorithm (Dynamic Load balancing). This Hybrid load balancing algorithm considers the server load before sending the request to the corresponding server. It oversees the previously selected server and this helps the selected server not participate in next server load decision. The sparse requests can be easily distributed among the n servers, more evenly.

### RL Algorithm

BEGIN PROCEDURE HYBRID_ALGO
SET static integer LAST_SELECT = 0;
SET array SERVERS = {s1, s2, s3……sn};
SET array SERVER_LOAD  = {l1,l2,l3……..ln};
WHILE (request) DO
    Integer POS = FIND_MIN_H (SERVER_LOAD, LAST_SELECT);
    LAST_SELECT = POS;
     GOTO SERVER [POS];
END WHILE
END PROCEDURE
BEGIN PROCEDURE FIND_MIN_H (SERVER_LOAD [1 to n] , LAST_SELECT)
    SET integer POS = 0;
    For I = 1 to n-1 do
    IF I=LAST_SELECT THEN
    Continue;
    END IF
    ELSE IF SERVER_LOAD [POS] > SERVER_LOAD[I] THEN
        POS = I;
    END ELSE IF
    RETURN POS;
END PROCEDURE

Sasikala and Ramesh, (2014) proposed an effective load balancing for cloud computing using hybrid AB algorithm. In order to reduce the waiting time and execution time, the hybrid AB algorithm is proposed. The hybrid AB algorithm is a combination of two dynamic algorithms, Ant Colony

Optimization and Bees Life algorithm (BLA). In this system design the user schedules the job to cloud through BLA and ant algorithm. Both give their calculation to control node. The BLA calculates the resource available in cloud and gives to control node. This ant colony checks for available resource in cloud by BLA result and makes perfect scheduling for user to cloud. The node is selected in cloud by its performance. If the node fails then another node is selected by which node having priority to become a master node. Thus, improves the performance of scheduling than another algorithm. This system achieves overload avoidance while the number of resources is increased, thus the resources

### AB Algorithm

**// pseudo code for Bees life algorithm**
Initialize population (*N* bees) at random
Evaluate fitness of population (fittest bee is the queen, *D* fittest following bees are drones, *W* fittest remaining bees areworkers)
*While* stopping criteria are not satisfied (Forming new population)
/* **reproduction behaviour***/
Generate *N* broods by **crossover** and **mutation**
Evaluate fitness of broods
If the fittest brood is fitter than the queen then replace the queen for the next generation
Choose *D* best bees among *D* fittest following broods and drones of current population (Forming next generationdrones)
Choose *W* best bees among *W* fittest remaining broods and workers of current population (to ensure food foraging)
/* **food foraging behaviour***/
Search of food source in *W* regions by *W* workers
Recruit bees for each region for neighbourhood search(more bees *(FBest)* for the best *B* regions and *(FOther)* forremaining regions)
Select the fittest bee from each region
Evaluate fitness of population (fittest bee is the queen, *D* fittest following bees are drones, *W* fittest remaining bees are workers)
End while

**//pseudo code for updation node**
If node has high priority than another node
Then
Choose the highest priority node as master node
Priority_option
Execution time storage capacity and performance level
If master node fails
Then select an alternative node which has next priority to become a master
If the other slave nodes fails
Then
Use a Gossip protocol to get a alive message from all nodes
Which node gives alive message
Then
Select all alive paths and find the shortest path by ant colony optimization algorithm
// pseudo code for ant colony optimization
Initialize parameters
Initialize pheromone trails
Create Ants
While stopping criteria is not reached do
Let all ants construct their solution
Update pheromone trails
Allow Daemon actions
End while

### CONCLUSION

Load balancing has been a major challenge in cloud computing. It provides proper resource utilization and improves the response time and cost that leads to customer satisfaction. Static algorithm uses the current state of node. It will not bother about the previous state of node. Dynamic algorithms use previous as well as current state of node to distribute the load. Research have shown that static and dynamic have not been fully efficient for load balancing. This gave room to hybrid algorithms. Hybrid methods inherit the properties from both static and dynamic load balancing techniques and attempts at overcoming the limitation of both algorithms. This paper is a study of various hybrid load balancing algorithms in cloud computing environment.